\documentclass[prd,nofootinbib,amsfonts,notitlepage,longbibliography]{revtex4-1}
\usepackage[utf8]{inputenc}
\usepackage[T1]{fontenc}
\usepackage{hyperref}
\usepackage{graphicx,bm,amsmath,color,scalerel}
\usepackage{bbold}
\usepackage{wasysym}
\usepackage{verbatim}

\newcommand{\be}{\begin{equation}}
\newcommand{\ee}{\end{equation}}
\newcommand{\beq}{\begin{eqnarray}}
\newcommand{\eeq}{\end{eqnarray}}
\newcommand{\ba}{\begin{align}}
\newcommand{\ea}{\end{align}}

\newcommand{\bigo}{\,\scaleobj{1.3}{\oplus}\,} %para hacer el \oplus grande, pero no tanto como el \bigoplus; el tamaño puede ajustarse. Hay que incluir en la definición los espacios que LaTeX pone automáticamente a los operadores, pero que no pone al definir este objeto nuevo

\begin{document}

\title{Spacetime from locality of interactions in deformations of special relativity: the example of $\kappa$-Poincaré Hopf algebra}
\author{J.M. Carmona}
\affiliation{Departamento de F\'{\i}sica Te\'orica,
Universidad de Zaragoza, Zaragoza 50009, Spain}
\author{J.L. Cort\'es}
\affiliation{Departamento de F\'{\i}sica Te\'orica,
Universidad de Zaragoza, Zaragoza 50009, Spain}
\author{J.J. Relancio}
\email{jcarmona@unizar.es, cortes@unizar.es, relancio@unizar.es}
\affiliation{Departamento de F\'{\i}sica Te\'orica,
Universidad de Zaragoza, Zaragoza 50009, Spain}

\begin{abstract}
A new proposal for the notion of spacetime in a relativistic generalization of special relativity based on a modification of the composition law of momenta is presented. Locality of interactions is the principle which defines the spacetime structure for a system of particles. The formulation based on $\kappa$-Poincaré Hopf algebra is shown to be contained in this framework as a particular example.   
\end{abstract}

\maketitle

\section{Introduction}

There is a consensus that the classical spacetime structure of special relativity (SR) will probably have to be modified in a quantum theory of gravity (see, for example, the different contributions in Ref.~\cite{Oriti:2009zz}). If this is the case, Poincaré invariance, the symmetry of spacetime at low energies, may have to be abandoned as a good symmetry of Nature at the energy scale at which effects of the nontrivial (`quantum') spacetime structure would start to emerge. An attractive idea is the possibility that the new physics beyond SR will still be compatible with the relativity principle, as happened in the past with the transition from Galilean physics to SR. In this sense, double (or deformed) special relativity (DSR) theories (see Ref.~\cite{AmelinoCamelia:2010pd} for a review), in which an energy scale $\Lambda$ is an observer-independent quantity, were proposed as relativistic generalizations of SR which could be relevant to account for such modifications appearing at this energy scale, even in the flat spacetime limit of a theory of quantum gravity~\cite{AmelinoCamelia:2011pe}.
 
Most of the attempts to consider a relativistic generalization of SR-kinematics, like DSR theories, take as a starting point a generalization of the dispersion relation, adding new $\Lambda$-dependent terms to the standard relation $E^2-p^2=m^2$. This requires a nonlinear ($\Lambda$-dependent) implementation of Lorentz transformations in momentum space in order to guarantee the invariance of the modified dispersion relation (MDR) imposed by the relativity principle. 

When one considers the energy-momentum conservation law in a process among particles then one finds that in a consistent relativistic kinematics the total four-momentum cannot be the sum of the four-momenta of the particles, since the linearity of the additive composition law of energy and momentum is not compatible with the nonlinearity introduced at the level of Lorentz transformations~\cite{AmelinoCamelia:2011yi}. Moreover, a modified composition law (MCL) requires to extend the nonlinear implementation of Lorentz transformations for one particle to a system of particles in a nontrivial way: since the nonlinear terms of the MCL mix the components of the momenta which are being composed, a boost transformation on these momenta will also generically mix them. In fact, once an expression for the total momentum of a system of particles in terms of the individual momenta (the MCL) has been chosen, it is possible to use the relativity principle to get the generalization of both the MDR of a particle and the Lorentz transformations of a system of particles~\cite{Carmona2012,Carmona2016b}. This suggests to take the composition law as the starting point of a generalized relativistic kinematics instead of the dispersion relation of particles.

Generalizations of SR based on modified dispersion relations and composition laws are naturally formulated in momentum space. On the other hand, a number of arguments identifying the existence of a minimum length (which is equivalent to an energy scale) as a qualitative feature of the quantum structure of spacetime~\cite{Garay1995} have led to the introduction of deformations in the Heisenberg algebra as a way to introduce such a scale. The first noncommutative model was proposed by Snyder~\cite{Snyder:1946qz} in connection with ultraviolet divergences in quantum field theory. Nonetheless, the study of noncommutativity was forgotten until this was considered as an approach to quantum gravity. In fact, as it is shown in Ref.~\cite{Hossenfelder:2012jw}, when one introduces gravity in the gedanken experiment of the Heisenberg microscope, nontrivial commutation relations in phase space appear. 

A number of works have explored the connection, through the product of plane waves, between a deformation of the Heisenberg algebra (phase-space commutation relations) and energy-momentum modified composition laws. In this context, the spacetime proposed by Snyder has got a MCL associated to it, as shown in Ref.~\cite{Battisti:2010sr}; in this case, and due to the fact that this is a covariant model, there is no modification in the dispersion relation nor in the Lorentz transformations. Another noncommutative spacetime model, $\kappa$-Minkowski, was inspired by the Hopf algebra known as $\kappa$-Poincaré~\cite{Majid:1995qg,Lukierski_pairing}, in which a MCL for the momenta, a modified Casimir (the MDR) and an implementation of nonlinear boosts in momentum space are chosen in order to satisfy the relativity principle. The modification in phase space appears as a consequence of this deformation, through the `pairing' operation~\cite{Kosinski_paring} making compatible the momentum space modification with the phase space one.

These mathematical approaches, however, lack a physical intuition about the connection between the noncommutative spacetime and the generalized kinematics given by the modified composition law. Our objective in this paper is to give such a physical interpretation in terms of the notion of locality in a classical picture of particle interactions.  

In SR, the locality of an interaction (crossing of worldlines) is an observer-independent fact. This assertion is however subtler in the case of a MCL, since the crossing of worldlines in spacetime coordinates which are canonically conjugated to the momentum coordinates is perceived differently according to the distance between the observer and the interaction point, a phenomenon named as relative locality, described systematically for the first time in Ref.~\cite{AmelinoCamelia:2011bm}. Such loss of the absolute notion of locality can be interpreted in geometrical terms, as a consequence of a curved momentum space~\cite{AmelinoCamelia:2011bm,AmelinoCamelia:2011pe}, but it can be easily understood from a variational principle~\cite{AmelinoCamelia:2011bm,Carmona:2011wc}. Heuristically, one has invariance under a transformation (translation) generated by the total momentum, which is conserved in the interaction. In the case of a MCL, this total momentum differs from the sum of the momentum variables of the particles. Then a translation is not a constant displacement, the same for the spacetime coordinates of all the particles. As a result, an interaction seen as local for one observer is no longer local for another observer related to the first one by a translation.

In this work we explore the possibility to look for a new spacetime, whose coordinates differ from the spacetime coordinates canonically conjugated to the momentum variables, such that absolute locality is compatible with the modified composition law. This will make possible to associate a spacetime with a given MCL that defines a generalization of SR. We will see that the relation between the $\kappa$-Poincaré Hopf algebra and $\kappa$-Minkowski spacetime through the pairing procedure is indeed of this type, but the idea of locality in the new spacetime gives rise to other possible extensions of the phase space of SR.

\section{Ingredients of a kinematics beyond special relativity}

\subsection{Relativistic generalization of special relativity kinematics through a modified composition law}
\label{sec:ingred1}

As we have argued in the Introduction, the composition law of momenta can be considered as the fundamental ingredient in a  relativistic generalization of SR-kinematics. We will review in this section how such a framework beyond SR can be constructed from a generic composition law. We follow the steps shown in Ref.~\cite{Carmona2016b}, considering as a starting point a covariant composition law\footnote{A generic covariant composition law cannot be constructed through a change of momentum variables.} and then obtaining any other MCL through a ``change of variables'' and a ``change of basis''. Let us first consider a generic covariant composition law, that is, a composition law which is invariant under the standard (linear) Lorentz transformations:
\be
\left(P \bigo Q\right)_\mu\,=\, P_\mu\,f_1\left(\frac{P^2}{\Lambda^2},\frac{P\cdot Q}{\Lambda^2},\frac{Q^2}{\Lambda^2}\right)+Q_\mu\,f_2\left(\frac{P^2}{\Lambda^2},\frac{P\cdot Q}{\Lambda^2},\frac{Q^2}{\Lambda^2}\right),
\ee
where we have defined $f_1$ and $f_2$ as functions whose arguments are Lorentz invariant quantities constructed from the four-vectors $P$ and $Q$, with the conditions:
\be
f_1\left(\frac{P^2}{\Lambda^2},0,0\right)\,=\,f_2\left(0,0,\frac{Q^2}{\Lambda^2}\right)\,=\,1,
\ee
so that the following consistency requirements are satisfied:
\be
\left(P \bigo Q\right)_\mu \left. \right|_{Q=0}\,=\, P_\mu \,,\quad \left(P \bigo Q\right)_\mu \left. \right|_{P=0}\,=\, Q_\mu.
\label{eq:consistency}
\ee

Now we will mix the momenta of the two-particle system through a ``change of variables'' 
\be
\left(P,Q\right) \rightarrow \left(\hat{p},\hat{q}\right) =\left(\mathcal{F}^L\left(P,Q\right),\mathcal{F}^R\left(P,Q\right) \right),
\ee
which has to satisfy the following properties:
\begin{align}
\mathcal{F}^L\left(P,0\right) = P \,,\quad \mathcal{F}^L\left(0,Q\right)\,=\,0\,, \quad & \mathcal{F}^R\left(0,Q\right)\,=\,Q\,, \quad   \mathcal{F}^R\left(P,0\right)\,=\,0 \,,
\label{eq:variables1} \\
P^2=(\mathcal{F}^L\left(P,Q)\right)^2 \,,\quad & Q^2=(\mathcal{F}^R\left(P,Q)\right)^2.
\label{eq:variables2}
\end{align}
Eq.~\eqref{eq:variables1} guarantees that when one of the momenta is equaled to zero, then the change of variables is just the identity function (the two-particle system reduces to a one-particle system), and Eq.~\eqref{eq:variables2} guarantees that there is no mixing of variables in the dispersion relations: momentum variables $\hat{p},\hat{q}$ satisfy standard dispersion relations. All the non-triviality is in the non-linear Lorentz transformations of the system of the two particles, and one can easily see~\cite{Carmona2016b} that the composition law $(\hat{p}\,\hat{\oplus}\, \hat{q})_\mu\equiv (P\bigo Q)_\mu$ satisfies consistency 
conditions analogous to Eq.~\eqref{eq:consistency} for the new momenta $\hat{p}$ and $\hat{q}$. This procedure allows one to define a  nonlinear composition law which is compatible with the standard dispersion relation through a proper modification of the transformation properties of the two-particle system (in the context of Hopf algebras, the $\{\hat{p}\}$ variables are usually known as the \emph{classical basis} of the algebra).

%These functions are generic functions depending on a fixed vector. 
Finally, one can get an arbitrary modified dispersion relation by introducing a ``change of basis'' 
\be
p_\mu\,=\,\mathcal{B}_\mu(\hat{p}),
\ee
where the $\mathcal{B}_\mu$ are four arbitrary $\Lambda$-dependent functions of a single momentum variable $\hat{p}$ such that $\lim_{\Lambda\to \infty}\mathcal{B}_\mu(\hat{p})=\hat{p}_\mu$. The composition law for the variables $p$ and $q$ is now~\cite{Carmona2016b}:
\begin{equation}
(p\oplus q)_\mu \equiv \mathcal{B}_\mu\left(\mathcal{B}^{-1}(p)\,\hat{\oplus}\,\mathcal{B}^{-1}(q)\right)\,.
\label{eq:MCLdef}
\end{equation}
From this change of basis one can then get the set of relations (also called `golden rules')~\cite{AmelinoCamelia:2011yi,Carmona2012,Carmona2016b} between a generic MCL and its corresponding MDR in a relativistic kinematics beyond special relativity.

\subsection{Relative locality}

A nonlinear composition law of momenta can be interpreted in terms of a nontrivial geometry of momentum space~\cite{AmelinoCamelia:2011bm,AmelinoCamelia:2011pe}, which modifies then the locality property of interactions. A modified conservation law implies that the sum of the momentum variables is not conserved: therefore, the system is no longer invariant under fixed translations of the canonically conjugate coordinates. Since the system has lost this symmetry property, local interactions for an observer (the crossing of worldlines) are seen as non-local for other observers.

This qualitative reasoning can be made quantitative by analyzing the interaction with a variational principle~\cite{AmelinoCamelia:2011bm,Carmona:2011wc}. Let us consider a process with $N$ interacting particles: we will number from $1$ to $N$ the incoming worldlines, and from $N+1$ to $2N$ the outgoing worldlines (since it is a classical interaction, the number and identity of particles is conserved in the process). The action is
\begin{equation}
S_\text{total}= S_\text{free}^\text{in}+S_\text{free}^\text{out} +S_\text{int}\,, 
\label{eq:action}
\end{equation}
where the free part of the action for the incoming worldlines is
\begin{equation}
S_\text{free}^\text{in}=\sum_{J=1}^{N}\int^{0}_{-\infty} ds \left(x^{\mu}_J \dot k^{J}_{\mu}+\mathcal{N}_J\left(C(k^{J})-m^2_J\right)\right)\,,
\end{equation}
and for the outgoing worldlines
\begin{equation}
S_\text{free}^\text{out}=\sum_{J=N+1}^{2N}\int_{0}^{\infty} ds \left(x^{\mu}_J \dot k^{J}_{\mu}+\mathcal{N}_J\left(C(k^{J})-m^2_J\right)\right)\,.
\end{equation}
In the above expressions $s$ is an arbitrary parametrization of the particle world line and $\mathcal{N}_J$ is the Lagrange multiplier imposing the mass shell condition 
\begin{equation}
C(k^J)=m^2_J\,,
\end{equation}
and the usual Poisson brackets are used
\begin{equation}
\left\lbrace k^{I}_{\nu}\,,\,x^{\mu}_J \right\rbrace =\delta^{\mu}_{\nu}\delta^I_J\,,
\end{equation}
that is, $x_J$ and $k^J$ are canonically conjugate variables.

The interaction contribution to the action is simply a Lagrange multiplier times the conservation law:
\begin{equation}
S_\text{int}=\left(\underset{N+1\leq J\leq 2N}{\bigoplus} k^J_\nu(0)\,\,-\underset{1\leq J\leq N}{\bigoplus} k^J_\nu(0)\right) \xi^\nu\,.
\end{equation}
The affine parameter ($s$) is chosen so that the interaction takes place at $s=0$ for every particle, and $\xi$ can be just considered to be a Lagrange multiplier to enforce the conservation of momentum at that point. 
After varying the action and integrating by parts one finds 
\begin{equation}
\delta S_\text{total}\,=\,\sum_J \int_{s_1}^{s_2}\left(\delta x^{\mu}_J \dot k^J_{\mu} - \delta k^{J}_{\mu}\left[\dot x^{\mu}_J-\mathcal{N}_J\frac{\partial C(k^J)}{\partial k^{J}_{\mu}} \right]\right)+\mathcal{R}\,,
\label{deltaS}
\end{equation}
where $\mathcal{R}$ contains both the result of varying $S_\text{int}$ and the boundary terms from the integration by parts, and $s_{1,2}$ are 0, $\infty$ or $-\infty$ depending on whether the corresponding term is incoming or outgoing. 
One gets
\be
\begin{split}
\mathcal{R}\, =\,\left(\underset{N+1\leq J\leq 2N}{\bigoplus} k^J_\nu(0)\,\,-\underset{1\leq J\leq N}{\bigoplus} k^J_\nu(0)\right) \delta\xi^\nu & +
\sum_{J=1}^{N} \left(x^{\mu}_J (0) -\xi^{\nu} \frac{\partial}{\partial k^J_{\mu}} \left[\underset{1\leq I\leq N}{\bigoplus} k^I_\nu\right](0)\right)\delta k^J_{\mu}(0) \\ & - 
\sum_{J=N+1}^{2N} \left(x^{\mu}_J (0) -\xi^{\nu} \frac{\partial}{\partial k^J_{\mu}} \left[\underset{N+1\leq I\leq 2N}{\bigoplus} k^I_\nu\right](0)\right) \delta k^J_{\mu}(0)\,,
\end{split}  
\ee
where the $x^{\mu}_J (0)$ are the spacetime coordinates of the ending (starting) point of the world line for $1\leq J\leq N$ ($N+1\leq J\leq 2N$).
According to the variational principle, the worldlines of particles should be such that $\delta S_\text{total}=0$ for any variation $\delta\xi^\mu$, $\delta x_J^\mu$, $\delta k^J_\mu$. From the vanishing term with $\delta \xi^\nu$ one obtains the conservation law for the momenta, and for the one with $\delta k^J_{\mu}(0)$ one sees that the equalities\footnote{The vanishing of the term proportional to $\delta x^\mu_J(s)$ in Eq.~(\ref{deltaS}) implies that the momenta are constant along each worldline.}
\begin{equation}
x^{\mu}_J (0)\,=\, \xi^{\nu} \frac{\partial}{\partial k^J_{\mu}} \left[\underset{1\leq I\leq N}{\bigoplus} k^I_\nu\right] \, \text{for } J=1,\ldots N
\,, \quad 
x^{\mu}_J (0)\,=\, \xi^{\nu} \frac{\partial}{\partial k^J_{\mu}} \left[\underset{N+1\leq I\leq 2N}{\bigoplus} k^I_\nu\right] \, \text{for } J=N+1,\ldots 2N \,,
\label{eq:endWL}
\end{equation}
have to be satisfied. 

The transformation
\begin{equation}
\delta \xi^\mu=a^\mu, \quad
\delta x^\mu_J=a^\nu\frac{\partial}{\partial k^J_{\mu}} \left[\underset{1\leq I\leq N}{\bigoplus} k^I_\nu\right] (J=1,\ldots N),\quad
 \delta x^\mu_J=a^\nu \frac{\partial}{\partial k^J_{\mu}} \left[\underset{N+1\leq I\leq 2N}{\bigoplus} k^I_\nu\right] (J=N+1,\ldots 2N),
\quad \delta k^J_\mu=0,
\label{eq:translation}
\end{equation}
connects different solutions from the variational principle. This is the translational invariance of the classical model. 
We see that the interaction will be local (all $x^\mu_J(0)$ coincide) only for the observer with $\xi^\mu=0$. This shows the relativity of locality. The purpose of this paper will be to find new spacetime coordinates such that an absolute locality of observers hold. 

\subsection{Noncommutative spacetime}

In many approaches to quantum gravity a noncommutative spacetime appears. This is associated to a minimum length as one cannot measure the position with more precision than the scale which deforms the standard commutators~\cite{Hossenfelder:2012jw}. This minimum length appears in different contexts, such as string theory~\cite{Garay1995} or DSR theories~\cite{Amelino-Camelia2001}.

Spacetime noncommutativity can be constructed by introducing new spacetime coordinates $\tilde{x}$ from canonical phase space coordinates ($x$, $p$):\footnote{For recent works using this construction see Ref.~\cite{Meljanac:2016jwk,*Loret:2016jrg,*Carmona:2017oit}.}
\be
\tilde{x}^\mu \,=\, x^\nu \,\varphi^\mu_\nu(p), \quad \quad \{p_\mu,x^\nu\}=\delta_\mu^\nu,\quad \{x^\mu,x^\nu\}=\{p_\mu,p_\nu\}=0,
\label{eq:NCspt}
\ee
where the function $\varphi^\mu_\nu(p)$ is really a function of $p/\Lambda$ by dimensional arguments, and has to satisfy that when $p/\Lambda \to 0$ it reduces to $\delta^\mu_\nu$. These new spacetime coordinates can be seen as a nontrivial subspace of four dimensions in a canonical phase space.

The Poisson brackets of the new spacetime coordinates in the one-particle system is 
\be
\{\tilde{x}^\mu, \tilde{x}^\sigma\} \,=\, \{ x^\nu \varphi^\mu_\nu(p), x^\rho \varphi^\sigma_\rho(p)\} \,=\, x^\nu \frac{\partial\varphi^\mu_\nu(p)}{\partial p_\rho} \,\varphi^\sigma_\rho(p) \,-\, x^\rho \frac{\partial\varphi^\sigma_\rho(p)}{\partial p_\nu} \,\varphi^\mu_\nu(p) \,=\, x^\nu \,\left(\frac{\partial\varphi^\mu_\nu(p)}{\partial p_\rho} \,\varphi^\sigma_\rho(p) \,-\, \frac{\partial\varphi^\sigma_\nu(p)}{\partial p_\rho} \,\varphi^\mu_\rho(p)\right).
\label{eq:commNCspt}
\ee
We will restrict ourselves in this work to the case of $\kappa$-Minkowski spacetime (see the next subsection) where the bracket turns out to be a combination of the $\tilde{x}$ coordinates with coefficients independent of $p$. The remaining phase space Poisson brackets are
\be
\{p_\nu, \tilde{x}^\mu\} \,=\,\varphi^\mu_\nu(p).
\ee
One can see that there are different representations for each noncommutative spacetime, i.e. different choices of $\varphi^\mu_\nu(p)$ leading to the same spacetime noncommutativity. In fact different choices of momentum variables in the canonical phase space lead to different representations of a noncommutative spacetime (see Appendix~\ref{ncst-rep}).

\subsection{The case of $\kappa$-Poincaré Hopf algebra}

The mathematical formalism of Hopf algebras offers a connection between two of the ingredients that we considered above: the presence of a modified composition law, which, in this context, is referred to as the ``coproduct'' operation, and a modified phase space, which is obtained from the coproduct through the mathematical procedure known as the ``pairing'' construction.

We will review these ideas in the specific case of the Hopf algebra known as $\kappa$-Poincaré, in the bicrossproduct basis~\cite{KowalskiGlikman:2002jr}. In this basis the coproduct of the generators of translations $P_\mu$ reads
\be
\Delta(P_0)\,=\,P_0\otimes \mathbb{1}+ \mathbb{1}\otimes P_0 \,,\qquad \Delta(P_i)\,=\,P_i\otimes \mathbb{1}  + e^{-P_0/\Lambda} \otimes P_i\,.
\label{coproduct}
\ee
The coproduct~(\ref{coproduct}) defines the composition law for the momenta
\be
(p\oplus q)_0\,=\,p_0 + q_0 \,, \qquad (p\oplus q)_i \,=\,p_i  + e^{-p_0/\Lambda} q_i\,.
\label{composition}
\ee
In Hopf algebras scenarios, given a coproduct, one can obtain the resultant Poisson brackets in phase space through the method of pairing. Following this algebraic procedure~\cite{KowalskiGlikman:2002jr,*Kowalski-Glikman2002} one obtains that the phase space Poisson brackets associated to the coproduct of momenta in the bicrossproduct basis of $\kappa$-Poincaré is  
\be
\lbrace\tilde{x}^0, \tilde{x}^i\rbrace \,=\,-\frac{\tilde{x}^i}{\Lambda}\,,\qquad  \lbrace\tilde{x}^0, p_0\rbrace \,=\,-1\,,\qquad  \lbrace\tilde{x}^0, p_i\rbrace \,=\,\frac{p_i}{\Lambda}\,,\qquad  \lbrace\tilde{x}^i, p_j\rbrace \,=\,-\delta^i_j\,,
\qquad \lbrace\tilde{x}^i, p_0\rbrace \,=\,0\,.
\ee

The first of these relations defines the spacetime noncommutativity known as $\kappa$-Minkowski spacetime. The previous commutation relations can be expressed in terms of the $\varphi(p)$ functions defined in the previous subsection, as
\be
\varphi^0_0(p)=1 \,,\qquad \varphi^0_i(p)=-\frac{p_i}{\Lambda} \,,\qquad \varphi^i_j(p)=\delta^i_j \,,\qquad \varphi^i_0(p)=0\,.
\label{eq:phibicross}
\ee
These are the functions $\varphi(p)$ that define the phase space of $\kappa$-Poincaré in the bicrossproduct basis. It is sometimes useful to use a covariant notation~\cite{Carmona2016b} and write Eq.~\eqref{eq:phibicross} in the more compact form:
\be
\varphi^\mu_\nu(p)\,=\,\delta^\mu_\nu-\frac{1}{\Lambda} n^\mu p_\nu + \frac{p\cdot n}{\Lambda} n^\mu n_\nu\,,
\label{phi}
\ee
where $n^\mu$ is a fixed vector of components $n^\mu=(1,0,0,0)$.

\section{Locality of interactions with a modified momentum composition law}

In the previous section we have explained how a modified composition law allows us to build a relativistic generalization of the kinematics of SR, with an associated modified dispersion relation and modified Lorentz transformations. We have seen that a MCL causes also a relativity of locality of interactions when using spacetime coordinates which are canonically conjugated to the momentum variables appearing in the MCL. We have also learned how a noncommutative spacetime can be derived from the introduction of new spacetime coordinates in a canonical phase space. We can ask the question: how is an interaction viewed in the noncommutative spacetime? Is there a choice of noncommutative spacetime coordinates such that interactions are still local for every observer in the generalized relativistic theory?

\subsection{Locality from a MCL: a first attempt}
\label{sec:firstattempt}

We consider the simplest process, an interaction with an initial state with two particles with momenta $k$, $l$ and a total momentum $k\oplus l$ and a final state of the two particles with momenta $p$, $q$ and total momentum $p\oplus q$. This is just a particular case ($N=2$) of the model of relative locality presented in the previous section. From Eq.~\eqref{eq:endWL} we get
\be
w^\mu(0) \,=\, \xi^\nu \frac{\partial(k\oplus l)_\nu}{\partial k_\mu}\,,\quad x^\mu(0) \,=\, \xi^\nu \frac{\partial(k\oplus l)_\nu}{\partial l_\mu}\,,\quad y^\mu(0) \,=\, \xi^\nu \frac{\partial(p\oplus q)_\nu}{\partial p_\mu} \,,\quad z^\mu(0) \,=\, \xi^\nu \frac{\partial(p\oplus q)_\nu}{\partial q_\mu}\,,
\ee
where $w^\mu(0)$, $x^\mu(0)$ are the spacetime coordinates of the end points of the worldlines of the particles in the initial state with momenta $k$, $l$ and $y^\mu(0)$, $z^\mu(0)$ the coordinates of the starting points of the worldlines of the particles in the final state with momenta $p$, $q$. 

In the case of the conventional composition law $p\oplus q = p + q$ one has a local interaction $w^\mu(0) = x^\mu(0) = y^\mu(0) = z^\mu(0) = \xi^\mu$, and the interaction of particles can be used to define events in spacetime. In the general case this is not possible any more. 

A first attempt to associate an event to the interaction is to introduce new spacetime coordinates $\tilde{x}$ from the canonical phase space coordinates ($x$, $p$), as in Eq.~\eqref{eq:NCspt}:
\be
\tilde{x}^\mu \,=\, x^\nu \,\varphi^\mu_\nu(p) ,
\label{eq:firstspt}
\ee
so that the end and starting points of the worldlines in this new spacetime are 
\begin{align}
& \tilde{w}^\mu(0) \,=\, \xi^\nu  \frac{\partial (k\oplus l)_\nu}{\partial k_\rho} \, \varphi^\mu_\rho(k) \,,\quad \tilde{x}^\mu(0) \,=\, \xi^\nu  \frac{\partial (k\oplus l)_\nu}{\partial l_\rho} \, \varphi^\mu_\rho(l) \,,
\nonumber \\ & \tilde{y}^\mu(0) \,=\, \xi^\nu  \frac{\partial (p\oplus q)_\nu}{\partial p_\rho} \, \varphi^\mu_\rho(p)\,,\quad \tilde{z}^\mu(0) \,=\, \xi^\nu \frac{\partial (p\oplus q)_\nu}{\partial q_\rho} \, \varphi^\mu_\rho(q) \,.
\end{align}
If it is possible to have a composition law and a set of functions $\varphi^\mu_\nu$ such that\footnote{Note that the conservation law of momenta implies that $k\oplus l = p\oplus q$.}
\be
\frac{\partial(k\oplus l)_\nu}{\partial k_\rho} \, \varphi^\mu_\rho(k) 
\,=\, \frac{\partial(k\oplus l)_\nu}{\partial l_\rho} \, \varphi^\mu_\rho(l) \,=\, \frac{\partial(p\oplus q)_\nu}{\partial p_\rho} \,\varphi^\mu_\rho(p) \,=\, \frac{\partial(p\oplus q)_\nu}{\partial q_\rho} \, \varphi^\mu_\rho(q) \,,
\label{loc0}
\ee
then one will have $\tilde{w}^\mu(0)=\tilde{x}^\mu(0)=\tilde{y}^\mu(0)=\tilde{z}^\mu(0)$ and the interaction will define an event in this new spacetime. If we consider the limit where one of the two momenta in the initial state goes to zero, using that $\lim_{l\to 0} (k\oplus l) = k$ [recall the consistency conditions Eq.~\eqref{eq:consistency}], and that in this limit the conservation law of momenta is just $k=p\oplus q$, Eq.~(\ref{loc0}) implies that
\be
\boxed{\varphi^\mu_\nu(p\oplus q) \,=\, \frac{\partial (p\oplus q)_\nu}{\partial p_\rho} \,\varphi^\mu_\rho(p) \,=\, \frac{\partial (p\oplus q)_\nu}{\partial q_\rho} \, \varphi^\mu_\rho(q)} \,.
\label{loc1}
\ee
If one takes the limit $p\to 0$ in each of the three expressions in the equality chain of Eq.~\eqref{loc1} then one has
\be
\varphi^\mu_\nu(q) \,=\, \lim_{p\to 0} \frac{\partial (p\oplus q)_\nu}{\partial p_\mu} \,=\, \varphi^\mu_\nu(q) \,,
\label{eq:limit1}
\ee
where we have used that $\lim_{p\to 0} \varphi^\mu_\rho(p) = \delta^\mu_\rho$ [recall the conditions over $\varphi^\mu_\nu(p)$ exposed after Eq.~\eqref{eq:NCspt}]. On the other hand, when one takes the limit $q\to 0$ one has
\be
\varphi^\mu_\nu(p) \,=\, \varphi^\mu_\nu(p) \,=\, \lim_{q\to 0} \frac{\partial (p\oplus q)_\nu}{\partial q_\mu} \,.
\label{eq:limit2}
\ee
Changing the labels $p$ and $q$ in Eq.~\eqref{eq:limit2} and comparing with Eq.~\eqref{eq:limit1}, we conclude that
\be
\lim_{p\to 0} \frac{\partial (p\oplus q)_\nu}{\partial p_\mu} \,=\,\lim_{p\to 0} \frac{\partial (q\oplus p)_\nu}{\partial p_\mu}\,.
\label{eq:limitsym}
\ee
This is a condition that not every modified composition law will satisfy. In particular, Eq.~\eqref{eq:limitsym} is satisfied if
\be
p\oplus q = q\oplus p\,,
\ee
that is, in the case of a commutative MCL. However, the case of a commutative MCL is a restriction that does not include the case of the $\kappa$-Poincaré Hopf algebra, see Eq.~\eqref{composition}.

Moreover, it can be proved that Eq.~\eqref{loc1} implies that the new spacetime $\tilde{x}$ is a commutative spacetime, which allows to identify new variables $\tilde{p}_\mu=g_\mu(p)$ such that $\{\tilde{p}_\nu, \tilde{x}^\mu\} = \delta^\mu_\nu$, and which compose additively, 
$[\tilde{p}\,\tilde{\oplus}\, \tilde{q}]_\mu = \tilde{p}_\mu + \tilde{q}_\mu$ (see Appendix~\ref{append-commut}). This means that the original variables $(x,p)$ are related by a canonical transformation to the variables of SR $(\tilde{x},\tilde{p})$, and the new spacetime that we have identified through the condition of locality is just the spacetime of SR. This identification will not work, however, for a generic (noncommutative) composition law (including the case of $\kappa$-Poincaré). We therefore need something beyond Eq.~\eqref{loc1}.  

\subsection{Second attempt: non-trivial implementation of spacetime in the two-particle system}
%two different spacetimes in the two-particle system}

As we mentioned in subsection~\ref{sec:ingred1}, in a relativistic generalization of SR based on a modified composition law, the extension of the one-particle system to the two-particle system is rather nontrivial. Since the composition of two momenta is a nonlinear function of each of them, the relativity principle requires that the Lorentz transformation of one of the momenta of the two-particle system depends also on the other momentum, in an order dependent way: each of the momenta transform differently under Lorentz transformations~\cite{Carmona2012}. This argument leads to consider the introduction of new spacetimes for the particles of a two-particle system different from the new spacetime of a single particle. The simplest way to do that is to consider
\be
\tilde{y}^\mu \,=\, y^\nu \,\varphi_{L\,\nu}^{\,\mu}(p, q) \,,\quad
\tilde{z}^\mu \,=\, z^\nu \, \varphi_{R\,\nu}^{\,\mu}(p, q) \,.
\ee  

The condition to have an event defined by the interaction is in this case
\be
\boxed{\varphi^\mu_\nu(p\oplus q) \,=\, \frac{\partial (p\oplus q)_\nu}{\partial p_\rho} \,\varphi_{L\,\rho}^{\,\mu}(p, q) \,=\, \frac{\partial (p\oplus q)_\nu}{\partial q_\rho} \, \varphi_{R\,\rho}^{\,\mu}(p, q)} \,.
\label{loc2}
\ee

In order to find the new spacetime for a two-particle system one has to introduce the functions $\phi_L$, $\phi_R$, defined by a composition law $p\oplus q$, through
\be
\phi_{L\,\sigma}^{\:\:\nu}(p, q) \,\frac{\partial(p\oplus q)_\nu}{\partial p_\rho} \,=\, \delta^\rho_\sigma\,,\quad
\phi_{R\,\sigma}^{\:\:\nu}(p, q) \,\frac{\partial(p\oplus q)_\nu}{\partial q_\rho} \,=\, \delta^\rho_\sigma \,.
\ee
These functions determine the spacetime of a two-particle system once the spacetime of a one-particle system (i.e. $\varphi$) has been fixed:
\be
\varphi_{L\,\sigma}^{\:\:\mu}(p, q) \,=\, \phi_{L\,\sigma}^{\:\:\nu}(p, q) \,\, \varphi^\mu_\nu(p\oplus q) \,,\quad
\varphi_{R\,\sigma}^{\:\:\mu}(p, q) \,=\, \phi_{R\,\sigma}^{\:\:\nu}(p, q) \,\, \varphi^\mu_\nu(p\oplus q)\,.
\label{phiL-phiR-phi}
\ee
Note that $\phi_{L\,\sigma}^{\:\:\nu}(p, 0) = \phi_{R\,\sigma}^{\:\:\nu}(0, q) = \delta^\nu_\sigma$, and then 
\be
\varphi_{L\,\sigma}^{\:\:\mu}(p, 0) = \varphi^\mu_\sigma(p)\,, \quad \varphi_{R\,\sigma}^{\:\:\mu}(0, q) = \varphi^\mu_\sigma(q)\,,
\ee
which is what one obtains directly by taking the limits $q\to 0$, $p\to 0$ of the relation Eq.~(\ref{loc2}).

The implementation of locality is compatible with an independent choice for a one-particle noncommutative spacetime (the $\varphi$ function) and a modified composition law. 
In contrast, these two ingredients are related when the composition law is defined through a product of plane waves in the noncommutative spacetime. Therefore, there is a much larger arbitrariness when one considers a relativistic generalization of SR based on the locality of the interactions than in the framework of Hopf algebras.
This suggests to look for additional criteria, beyond locality, that a generalization of SR should satisfy. The algebraic structure of the spacetime of the two-particle system could be the place to introduce new restrictions on a relativistic generalization of SR. Results along this line will be presented elsewhere~\cite{Carmona:2017a}.

In order to show how the framework of Hopf algebras can be interpreted in terms of the locality of interactions we will consider the simplest way to restrict the implementation of locality establishing a relation between the one-particle spacetime coordinates and the modification of the composition law of momenta. It corresponds to an intermediate step between the first (failed) attempt and the second attempt. In the first attempt we considered an introduction of the new spacetime independently for each particle while in the second attempt the new spacetime coordinates of each particle depend on the momenta of both particles. In the intermediate step the spacetime of one of the particles is independent of the other particle while the spacetime coordinates of the other particle depend on the momenta of both particles. Then one has
\be
\varphi^{\:\:\mu}_{L\rho}(p, q) \,=\, \varphi^{\:\:\mu}_{L\rho}(p, 0) \,=\,\varphi^\mu_\rho(p)\,,
\ee
and Eq.~(\ref{loc2}) implies that
\be
\varphi^\mu_\nu(p\oplus q) \,=\, \frac{\partial (p\oplus q)_\nu}{\partial p_\rho} \,\varphi^\mu_\rho(p)\,,
\label{varphi-oplus}
\ee
which could be used to determine the composition law of momenta for a given one-particle noncommutative spacetime (i.e., for a given function $\varphi$). Taking the limit $p\to 0$ one has
\be
\varphi^\mu_\nu(q)  \,=\, \lim_{p\to 0}  \frac{\partial (p\oplus q)_\nu}{\partial p_\mu}\,,
\label{magic_formula}
\ee
which gives the one-particle spacetime corresponding to a certain composition law. Eq.~(\ref{magic_formula}) has a simple interpretation: the change of the momentum variable $p_\mu$ by an infinitesimal transformation with parameters $\epsilon$ generated by the noncommutative spacetime coordinates $\tilde{x}$ is
\be
\delta p_\mu \,=\, \epsilon_\nu \lbrace\tilde{x}^\nu, p_\mu\rbrace \,=\, - \epsilon_\nu \varphi^\nu_\mu(p) \,=\, - \epsilon_\nu \lim_{q\to 0} \frac{\partial(q\oplus p)_\mu}{\partial q_\nu} \,=\, - \left[(\epsilon\oplus p)_\mu - p_\mu\right] \,.
\label{deltap}
\ee
The transformation generated by the noncommutative spacetime coordinates is a displacement in momentum space defined by the modified momentum composition law. 

Similar results would have been obtained if we would have considered the case where it is the spacetime of the particle with momentum $q$ which is chosen to be independent of the particle with momentum $p$. In this case one would have
\be
\varphi^{\:\:\mu}_{R\rho}(p, q) \,=\, \varphi^{\:\:\mu}_{R\rho}(0, q) \,=\,\varphi^\mu_\rho(q) \,,
\ee
\be
\varphi^\mu_\nu(p\oplus q) \,=\, \frac{\partial (p\oplus q)_\nu}{\partial q_\rho} \,\varphi^\mu_\rho(q) \,,
\ee
\be
\varphi^\mu_\nu(p)  \,=\, \lim_{q\to 0}  \frac{\partial (p\oplus q)_\nu}{\partial q_\mu}\,,
\ee
and
\be
\delta p_\mu \,=\, - \left[(p\oplus \epsilon)_\mu - p_\mu\right] \,.
\ee

\subsection{Application to $\kappa$-Poincaré}

One can wonder about the relation between the relativistic generalization of SR based on the identification of new spacetime coordinates for a two-particle system such that interactions are local, and the generalization based on the use of Hopf algebras. In the framework of Hopf algebras a generalization of SR is defined by a deformed Poincaré algebra, a coproduct of momenta, and a coproduct of Lorentz generators. From a kinematical perspective, the deformed Poincaré algebra defines a modified dispersion relation through the Casimir of the deformed algebra, the coproduct of momentum defines a composition law of momenta, and the coproduct of Lorentz generators defines the Lorentz transformations of the momentum variables of a two-particle system. On the other hand, the change of basis and change of variables that applied to a covariant composition law allows one to define a modified composition law can be used to identify the modified dispersion relation (through the change of basis) and the modified Lorentz transformation of two momentum variables (through the change of variables and change of basis), which guarantees that the Lorentz transformation of the composition of two momenta is the composition of the transformed momenta (relativity principle). Since the Hopf algebra framework implements the relativistic invariance then it will be possible to associate a covariant composition law, a change of basis and a change of variables to a given Hopf algebra. 

What is the relationship between the noncommutative spacetime coordinates for which the interaction in the relativistic generalization of SR is local and the introduction of spacetime coordinates through the mechanism of pairing in Hopf algebras?

We will consider the case of $\kappa$-Poincaré Hopf algebra which corresponds to $\kappa$-Minkowski noncommutative spacetime
\be
\lbrace\tilde{x}^\mu, \tilde{x}^\nu\rbrace \,=\, \frac{1}{\Lambda} \,\left( \tilde{x}^\mu n^\nu -  \tilde{x}^\nu n^\mu \right) \,,
\ee
so that $\varphi^\mu_\nu(k)$ is such that
\be
\frac{\partial\varphi^\mu_\alpha(k)}{\partial k_\beta} \varphi^\nu_\beta(k) - \frac{\partial\varphi^\nu_\alpha(k)}{\partial k_\beta} \varphi^\mu_\beta(k) \,=\, \frac{1}{\Lambda} \,\left( \varphi^\mu_\alpha(k) n^\nu - \varphi^\nu_\alpha(k) n^\mu \right)\,.
\ee

For simplicity, we will take the function $\varphi^\mu_\nu(k)$ of the bicrossproduct basis, Eq.~\eqref{phi}. If now we impose that $\varphi_{L\,\nu}^{\:\:\mu}(p, q)\,=\, \varphi^\mu_\nu(p)$, we can determine unequivocally the composition law from Eq.~(\ref{varphi-oplus}). The result (see Appendix~\ref{append-bicross}) is just the momentum composition law corresponding to the coproduct of the momentum in the bicrossproduct basis. This is a proof that the framework of Hopf algebras to study relativistic theories in a noncommutative spacetime is contained in a general study of relativistic theories based on the implementation of locality.  

To complete the proof one can show explicitly (see Appendix~\ref{append-2pLT}) how the modified Casimir of the deformed Poincaré algebra and the coproduct of Lorentz generators in the bicrossproduct basis are derived from the corresponding function $\varphi^\mu_\nu(k)$ and the composition law derived from it through the particular implementation of locality with $\varphi_{L\,\nu}^{\:\:\mu}(p, q)\,=\, \varphi^\mu_\nu(p)$.    

To end up the discussion of the $\kappa$-Poincaré algebra from the perspective of locality of interactions we can determine $\varphi_{R\,\nu}^{\:\:\mu}(p, q)$ through Eq.~\eqref{phiL-phiR-phi}:
\be
\varphi_{R\,\nu}^{\:\:\mu}(p, q)\,=\,\delta^\mu_\nu \,e^{pn/\Lambda}+\frac{1}{\Lambda}n^\mu\left(n_\nu(e^{pn/\Lambda}\,pn+qn+(1-e^{pn/\Lambda})\,\Lambda)-e^{pn/\Lambda}\,p_\nu-q_\nu\right) ,
\ee
or, in components,
\be
\varphi_{R\,0}^{\:\:0}(p, q)\,=\,1 \,,\qquad \varphi_{R\,0}^{\:\:i}(p, q)\,=\,0 \,,\qquad \varphi_{R\,i}^{\:\:0}(p, q)\,=\,-\frac{e^{p_0/\Lambda}p_i+q_i}{\Lambda}\,, \qquad \varphi_{R\,j}^{\:\:i}(p, q)\,=\,e^{p_0/\Lambda}\delta^i_j \,.
\ee
We notice that, as expected,
\be
\varphi_{R\,\nu}^{\:\:\mu}(0, q)\,=\, \varphi^\mu_\nu(q) .
\ee
From this, one obtains the two-particle phase-space Poisson brackets that are different from zero:
\be
\begin{split}
\lbrace\tilde{y}^0, \tilde{y}^i\rbrace \,=\,-\frac{\tilde{y}^i}{\Lambda}\,,\qquad  \lbrace\tilde{y}^0, p_0\rbrace \,=\,-1\,,\qquad  \lbrace\tilde{y}^0, p_i\rbrace \,=\,\frac{p_i}{\Lambda}\,,\qquad  \lbrace\tilde{y}^i, p_j\rbrace \,=\,-\delta^i_j\,, \qquad \lbrace\tilde{y}^0, \tilde{z}^i\rbrace \,=\,-\frac{\tilde{z}^i}{\Lambda},\\
\lbrace\tilde{z}^0, \tilde{z}^i\rbrace \,=\,-\frac{\tilde{z}^i}{\Lambda}\,,\qquad  \lbrace\tilde{z}^0, q_0\rbrace \,=\,-1\,,\qquad  \lbrace\tilde{z}^0, q_i\rbrace \,=\,\frac{e^{p_0/\Lambda}p_i+q_i}{\Lambda}\,,\qquad  \lbrace\tilde{z}^i, q_j\rbrace \,=\,-e^{p_0/\Lambda}\delta^i_j\,.
\end{split}
\ee
It is remarkable  that all the Poisson brackets of spacetime coordinates are independent of momenta. 

\section{Outlook}

In this work we have considered relativistic generalizations of SR which are implemented through a nonadditive composition law for the momenta. We have seen how this ingredient is enough to define modified dispersion relations and modified Lorentz transformations in an unequivocal way. In particular, DSR theories are examples of this kind of generalization of SR.

We have also seen that a modified composition law of momenta is related with the loss of the absolute notion of locality in a classical model of interactions (there is no crossing of worldlines for every observer). In this case, a `relative locality' appears, in which distant observers must describe the interaction in a complicated way in the ordinary spacetime coordinates, canonically conjugated to the momentum variables. In particular, translational invariance is implemented in a nontrivial way in such a spacetime. This might represent a difficulty in the construction of a quantum field theory version of these type of theories.

Our main result has been to show that it is possible to define a noncommutative spacetime for particles participating in an interaction in such a way that the interaction is seen as local for every observer. The absolute locality of interactions is then implemented by Eq.~\eqref{loc2}. In doing so, the noncommutative spacetime of the one-particle system, given by the function $\varphi^\mu_\nu(p)$ which defines the new coordinates $\tilde{x}^\mu$, needs a nontrivial extension to the spacetime of the two-particle system, given by the functions $\varphi_{L\nu}^{\:\:\mu}$ and $\varphi_{R\nu}^{\:\:\mu}$, which define the coordinates of the two particles, $\tilde{y}^\mu$ and $\tilde{z}^\mu$. Both functions, $\varphi_{L\nu}^{\:\:\mu}$ and $\varphi_{R\nu}^{\:\:\mu}$, can be obtained from the MCL and the function $\varphi^\mu_\nu(p)$ through Eq.~\eqref{loc2}. 

In the condition implementing locality, Eq.~\eqref{loc2}, the function $\varphi^\mu_\nu(p)$ defining the noncommutative spacetime of a single particle appears as an input ingredient, independent of the MCL defining the relativistic generalization of SR. There exists then a large freedom to introduce a noncommutative spacetime in a relativistic theory beyond SR in a way compatible with the locality of interactions. An interesting particular case is the one in which the new spacetime of the two-particle system is such that the coordinates of one of the particles depend only on its own momentum. In this particular case there is a one-to-one correspondence between the MCL and the function $\varphi^\mu_\nu(p)$ [see Eqs.~\eqref{magic_formula}-\eqref{varphi-oplus}], and the coordinates of the one-particle system (limit of the two-particle system when one of the momenta goes to zero) are the generators of ``deformed'' translations Eq.~\eqref{deltap} in momentum space.

We can do several remarks that derive from the present work. Firstly, 
this framework provides a different perspective of the idea of noncommutative coordinates, which has been widely discussed in different contexts, particularly in connection with a quantum spacetime structure that could be of relevance for the quantum gravity problem. In our proposal, a noncommutative spacetime emerges in fact from a locality condition in a classical model which generalizes SR instead of from the implementation of a possible minimal length in a quantum spacetime. 

Secondly, the new perspective of the origin of the noncommutativity of spacetime could give a new way to explore the possibility to go beyond relativistic quantum field theory based on the incorporation in the framework of field theory of a modified composition law of momenta together with some notion of locality.   

Thirdly, we gave a physical interpretation (based on the introduction of a new spacetime to make a modified composition law of momenta compatible with locality) to the formal, mathematical procedure of ``pairing'', which is used in the context of Hopf algebras to obtain a noncommutative spacetime (and phase space) associated to the coproduct of the coalgebra (which is in fact the MCL in our language). The identification of the mathematical formalism of Hopf algebras as a particular case of a formulation of a relativistic generalization of SR based on locality opens up the possibility to go beyond this formalism~\cite{Carmona:2017a} in the study of relativistic theories with a noncommutative spacetime.

\section*{Acknowledgments}
This work is supported by the Spanish MINECO FPA2015-65745-P (MINECO/FEDER) and Spanish DGIID-DGA Grant No. 2015-E24/2. We acknowledge useful discussions with Niccolò Loret and the participation of Lorenzo Piga in the first stages of this work.

\appendix

\section{Different representations of a noncommutative spacetime}
\label{ncst-rep}

One can consider a canonical transformation in phase space $(x, p) \to (x', p')$ 
\be
p_\mu \,=\, f_\mu(p')\,,\quad x^\mu \,=\, x'^\nu g^\mu_\nu(p')\,,
\ee
for any non linear change of momentum variables, i.e, for any set of functions $f_\mu$ of the momentum variables, with
\be
g^\mu_\rho(p') \frac{\partial f_\nu(p')}{\partial p'_\rho} \,=\, \delta^\mu_\nu \,.
\ee
The canonical transformation can be used to write the noncommutative spacetime coordinates in terms of the new canonical phase space coordinates
\be
\tilde{x}^\mu \doteq x^\nu \varphi^\mu_\nu(p) \,=\, x'^\rho g^\nu_\rho(p') \varphi^\mu_\nu(f(p')) \,.
\ee
If one introduces
\be
\varphi'^\mu_\rho(p') \doteq g^\nu_\rho(p') \varphi^\mu_\nu(f(p')) \,=\, \frac{\partial p'_\rho}{\partial p_\nu} \varphi^\mu_\nu(p)\,,
\ee
the noncommutative spacetime coordinates are given by
\be
\tilde{x}^\mu \,=\, x'^\rho \varphi'^\mu_\rho(p') \,.
\ee
One has then a different representation of a noncommutative spacetime, i.e. a different set of functions $\varphi^\mu_\nu$ which allow to express the noncommutative spacetime coordinates in terms of canonical phase space coordinates, for each choice of momentum variables in phase space. 

\section{SR spacetime from the locality condition for a commutative MCL}
\label{append-commut}

We saw in subsection~\ref{sec:firstattempt} that the first attempt to implement locality through the condition Eq.~\eqref{loc1} is not valid for a generic composition law, since it implies Eq.~\eqref{eq:limitsym}. As we mentioned there, in those cases in which it is valid (such as for a commutative MCL), it can be proved that the spacetime defined by Eq.~\eqref{eq:firstspt} is the one of SR. Let us first see that the new spacetime coordinates $\tilde{x}^\mu$ are indeed commutative coordinates. Deriving with respect to $p_\sigma$ both sides of the first equality in Eq.~\eqref{loc1}, we get
\be
\frac{\partial \varphi^\mu_\nu(p\oplus q)}{\partial p_\sigma} \,=\, \frac{\partial}{\partial q_\rho} \left(\frac{\partial [p\oplus q]_\nu}{\partial p_\sigma}\right) \,\varphi^\mu_\rho(q) \,.
\label{eq:pre1}
\ee
If we use that the left hand side of the previous equation can be written as (applying the chain rule)
\be
\frac{\partial \varphi^\mu_\nu(p\oplus q)}{\partial p_\sigma} \,=\, \frac{\partial \varphi^\mu_\nu(p\oplus q)}{\partial [p\oplus q]_\rho} \,\frac{\partial [p\oplus q]_\rho}{\partial p_\sigma}\,,
\label{eq:pre2}
\ee
and take the limit $p\to 0$ of the right hand sides of Eqs.~\eqref{eq:pre1} and Eq.~\eqref{eq:pre2}, we get, using also Eq.~\eqref{eq:limit2},
\be
 \frac{\partial \varphi^\mu_\nu(q)}{\partial q_\rho} \,\varphi^\sigma_\rho(q) \,=\,  \frac{\partial \varphi^\sigma_\nu(q)}{\partial q_\rho} \,\varphi^\mu_\rho(q) \,.
\label{eq:pre3}
\ee
Comparing Eq.~\eqref{eq:pre3} with Eq.~\eqref{eq:commNCspt}, we see that, indeed, $\{\tilde{x}^\mu, \tilde{x}^\sigma\}=0.$

The commutativity of $\tilde{x}$ allows one to identify new momentum variables $\tilde{p}_\mu=g_\mu(p)$ such that $(\tilde{x},\tilde{p})$ are canonically conjugate variables, as it was the case for $(x,p)$, that is: there is a canonical transformation relating both sets of variables. One can then show that the new momentum variables compose additively, that is,
\be
[\tilde{p}\,\tilde{\oplus} \,\tilde{q}]_\mu \,\doteq\, g_\mu(p\oplus q) \,=\, \tilde{p}_\mu + \tilde{q}_\mu\,,
\label{lcl}
\ee
[where the composition law $\tilde{\oplus}$ has been defined as in Eq.~\eqref{eq:MCLdef}], so that $(\tilde{x},\tilde{p})$ is indeed the phase space of SR.

In order to prove this, let us first show that Eq.~\eqref{loc1} is invariant under changes of momentum variables. Considering new momentum variables $\tilde{p}_\mu = g_\mu(p)$, the first factor in the second term in Eq.~\eqref{loc1} is 
\begin{equation}
\frac{\partial[\tilde{p}\,\tilde{\oplus}\, \tilde{q}]_\nu}{\partial \tilde{p}_\rho} \,=\, \frac{\partial g_\nu (p\oplus q)}{\partial g_\rho (p)}\,=\,\frac{\partial g_\nu (p\oplus q)}{\partial [p\oplus q]_\lambda}\,\frac{\partial [p\oplus q]_\lambda}{\partial p_\sigma }\,\frac{\partial p_\sigma}{\partial g_\rho (p)}\,,
\end{equation}
and the second factor  
\begin{equation}
\tilde{\varphi}^\mu _\rho (\tilde{p}) \,= \, \lim\limits_{g(k)\rightarrow 0}\,\frac{\partial g_\rho (k\oplus p)}{\partial g_\mu (k)}\,=\,\lim\limits_{k\rightarrow 0}\,\frac{\partial g_\rho (k\oplus p)}{\partial [k\oplus p]_\xi}\,\frac{\partial [k\oplus p]_\xi }{\partial k_\eta}\,\frac{\partial k_\eta}{\partial g_\mu (k)}\,=\,\,\frac{\partial g_\rho (p)}{\partial p_\xi}\,\varphi _\xi ^\mu (p)\,,
\label{phi_transformation}
\end{equation}
where we used  Eq.~\eqref{eq:limit2} and the fact that the change of variables is such that $\tilde{p}=0\implies p=0$. So putting all this together one finds that the second term of Eq.~\eqref{loc1} transforms as 
\begin{equation}
\frac{\partial[\tilde{p}\,\tilde{\oplus}\, \tilde{q}]_\nu}{\partial \tilde{p}_\rho} \,\tilde{\varphi}^\mu _\rho (\tilde{p}) \,= \, \frac{\partial g_\nu (p\oplus q)}{\partial [p\oplus q]_\lambda} \,\frac{\partial[p\oplus q]_\lambda}{\partial p_\sigma} \,\varphi^\mu _\sigma (p) \,.
\end{equation}
The transformation of the third term of Eq.~\eqref{loc1} can be obtained following the same steps 
\begin{equation}
\frac{\partial[\tilde{p}\,\tilde{\oplus}\, \tilde{q}]_\nu}{\partial \tilde{q}_\rho} \,\tilde{\varphi}^\mu _\rho (\tilde{q}) \,= \, \frac{\partial g_\nu (p\oplus q)}{\partial [p\oplus q]_\lambda} \,\frac{\partial[p\oplus q]_\lambda}{\partial q_\sigma} \,\varphi^\mu _\sigma (q) \,.
\end{equation}
Finally, the transformation of the first term of Eq.~\eqref{loc1} follows from Eq.~\eqref{phi_transformation}, giving 
\begin{equation}
\tilde{\varphi}^\mu _\nu (\tilde{p}\,\tilde{\oplus}\, \tilde{q}) \,= \,\frac{\partial g_\nu (p\oplus q)}{\partial [p\oplus q]_\lambda}\,\varphi _\lambda ^\mu (p\oplus q)\,.
\end{equation}

In conclusion, we see that if Eq.~\eqref{loc1} is satisfied for the variables ($p$, $q$), it will also be satisfied for the variables ($\tilde{p}$, $\tilde{q}$) obtained by a change of momentum variables $\tilde{p}_\mu = g_\mu(p)$, $\tilde{q}_\mu = g_\mu(q)$. This means that one has Eq.~\eqref{loc1} for the new momentum variables and its corresponding composition law, with $\tilde{\varphi}^\mu_\nu(\tilde{k}) = \delta^\mu_\nu$ since one has locality in their canonical spacetime coordinates. But in this case one has 
\be
\frac{\partial[\tilde{p}\,\tilde{\oplus}\, \tilde{q}]_\nu}{\partial \tilde{p}_\mu} \,=\,  \frac{\partial[\tilde{p}\,\tilde{\oplus}\, \tilde{q}]_\nu}{\partial \tilde{q}_\mu} \,=\, \delta^\mu_\nu\,,
\ee
and then $[\tilde{p}\,\tilde{\oplus}\, \tilde{q}]_\nu = \tilde{p}_\nu + \tilde{q}_\nu$, as we stated in Sect.~\ref{sec:firstattempt}.

\section{MCL of $\kappa$-Poincaré in the bicrossproduct basis through locality}
\label{append-bicross}

In this appendix we will see how the composition law is obtained unequivocally through the locality condition when  $\varphi_{L\,\nu}^{\:\:\mu}(p, q)\,=\, \varphi^\mu_\nu(p)$ taking the $\varphi^\mu_\nu(p)$ of Eq.~\eqref{eq:phibicross} corresponding to  $\kappa$-Poincaré in the bicrossproduct basis. From Eq.~\eqref{varphi-oplus}, the system of equations that we have to solve is
\be
\begin{split}
\text{For}\quad \mu\,&=\,0,\quad \nu\,=\,0 \qquad 1\,=\,\frac{\partial (p\oplus q)_0}{\partial p_0}-\frac{\partial (p\oplus q)_0}{\partial p_i}\,\frac{p_i}{\Lambda} \,.\\
\text{For}\quad \mu\,&=\,i,\quad \nu\,=\,0 \qquad 0\,=\,\frac{\partial (p\oplus q)_0}{\partial p_j}\,\delta^i_j \,.\\
\text{For}\quad \mu\,&=\,0,\quad \nu\,=\,i \qquad -\frac{(p \oplus q)_i}{\Lambda}\,=\,\frac{\partial (p\oplus q)_i}{\partial p_0}-\frac{\partial (p\oplus q)_i}{\partial p_j}\,\frac{p_j}{\Lambda}\,.\\
\text{For}\quad \mu\,&=\,i,\quad \nu\,=\,j \qquad \delta^i_j\,=\,\frac{\partial (p\oplus q)_j}{\partial p_i}\,.\\
\end{split}
\ee 
From the second equation we can see that the zero component of the composition as a function of the momentum $p$ depends only on its zero component, but the first equation tells us that $(p\oplus q)_0$ is linear in $p_0$. Hence, the condition Eq.~\eqref{eq:consistency} gives the result 
\be
(p\oplus q)_0\,=\,p_0+q_0\,.
\ee
From the fourth equation we see that the spatial component of the composition is linear in the spatial component of the first momentum, giving
\be (p\oplus q)_i\,=\,p_i+q_i f(p_0,q_0,\vec{q}^2)\,.
\ee
With this form of the composition law we can solve the differential equation appearing in the third equation, giving
\be
f\,=\,e^{-p_0/\Lambda}\,g(q_0,\vec{q}^2)\,,
\ee
but in order to satisfy Eq.~\eqref{eq:consistency}, $g(q_0,\vec{q}^2)\,=\,1$, so the composition finally reads
\be
(p\oplus q)_0\,=\,p_0+q_0\,,\qquad (p\oplus q)_i\,=\,p_i+q_i e^{-p_0/\Lambda}\,.
\label{mcl-bcb}
\ee

\section{Lorentz transformation in the bicrossproduct basis from locality}
\label{append-2pLT}

We start from the ten-dimensional Lie algebra including the Lorentz algebra and the $\kappa$-Minkowski noncommutative spacetime coordinates as generators. The noncommutativity of spacetime requires a modification of the Poisson brackets of boost generators $J^{0i}$ and spacetime coordinates $\tilde{x}^\mu$:
\be
\lbrace \tilde{x}^0, J^{0i} \rbrace \,=\, \tilde{x}^i + \frac{1}{\Lambda} J^{0i} \,,\qquad \lbrace \tilde{x}^j, J^{0i}\rbrace \,=\, \delta^i_j \tilde{x}^0 + \frac{1}{\Lambda} J^{ji}\,.
\ee
Using the notation
\be
\tilde{x}^\mu \,=\, x^\nu \varphi^\mu_\nu(p) \,,\quad J^{0i} \,=\, x^\mu \psi^{0i}_\mu(p) \,, \quad J^{ij} \,=\, x^\mu \psi^{ij}_\mu (p) = x^j p_i - x^i p_j\,,
\ee
the modified Poisson brakets lead to the system of equations
\be
\frac{\partial\varphi^0_\mu}{\partial p_\nu} \psi^{0i}_\nu - \frac{\partial\psi^{0i}_\mu}{\partial p_\nu} \varphi^0_\nu \,=\, \varphi^i_\mu + \frac{1}{\Lambda} \psi^{0i}_\mu \,,\qquad
\frac{\partial\varphi^j_\mu}{\partial p_\nu} \psi^{0i}_\nu - \frac{\partial\psi^{0i}_\mu}{\partial p_\nu} \varphi^j_\nu \,=\, \delta^i_j \varphi^0_\mu + \frac{1}{\Lambda} \left(\delta^i_\mu p_j - \delta^j_\mu p_i\right)\,.
\label{J0i-xtilde}
\ee
If we insert $\varphi^\mu_\nu(p)$ of Eq.~\eqref{eq:phibicross} corresponding to  $\kappa$-Poincaré in the bicrossproduct basis we can solve (\ref{J0i-xtilde}) for $\psi^{0i}_\mu$. The result is
\be
\psi^{0i}_0 \,=\, - p_i \,,\qquad \psi^{0i}_j \,=\, \delta^i_j \,\frac{\Lambda}{2} \left[e^{-2 p_0/\Lambda} - 1 - \frac{\vec{p}^2}{\Lambda^2}\right] + \,\frac{p_i p_j}{\Lambda}\,.
\label{psi(0i)}
\ee
From this result we can calculate the Poisson brackets of $p_\mu$ and $J^{0i}$
\be
\lbrace p_0, J^{0i}\rbrace \,=\, - p_i \,,\qquad \lbrace p_j, J^{0i}\rbrace  \,=\, \delta^i_j \,\frac{\Lambda}{2} \left[e^{-2 p_0/\Lambda} - 1 - \frac{\vec{p}^2}{\Lambda^2}\right] + \,\frac{p_i p_j}{\Lambda} \,,
\ee
to be compared with the deformed Poincar\'e algebra in  the bicrossproduct basis~\cite{KowalskiGlikman:2002jr}
\be
[P_0, N_i] \,=\, - i\, P_i    \,,\qquad [P_j, N_i] \,=\, i\, \delta^i_j \,\frac{\Lambda}{2} \left[e^{-2 P_0/\Lambda} - 1 - \frac{\vec{P}^2}{\Lambda^2}\right] + \,i \,\frac{P_i P_j}{\Lambda} \,.
\ee
We can also identify the function of momenta $C(p_0, \vec{p}^2)$ which is invariant under boosts. It has to satisfy the equation
\be
0 \,=\, \lbrace C, J^{0i}\rbrace \,=\, \frac{\partial C}{\partial p_0} \lbrace p_0, J^{0i}\rbrace + 2 p_j \frac{\partial C}{\partial \vec{p}^2} \lbrace p_j, J^{0i}\rbrace \,=\, - p_i \,\frac{\partial C}{\partial p_0} + p_i \Lambda \left[e^{-2 p_0/\Lambda} - 1 + \frac{\vec{p}^2}{\Lambda^2}\right] \,\frac{\partial C}{\partial \vec{p}^2}\,.
\ee
The solution for $C$, such that $C\approx p_0^2 - \vec{p}^2$ when $(|p_\mu|/\Lambda) \ll 1$, is
\be
C(p_0, \vec{p}^2) \,=\, \Lambda^2 \left(e^{p_0/\Lambda} + e^{-p_0/\Lambda} - 2\right) - e^{p_0/\Lambda} \vec{p}^2 \,,
\ee
to be compared with  the Casimir of the deformed Poincare algebra in the bicrossproduct basis\footnote{The correspondence between the expressions for $\lbrace p_\mu, J^{0i}\rbrace$ and $[P_\mu, N_i]$ guarantees that one will find the same function $C$.}~\cite{KowalskiGlikman:2002jr}
\be
C(P_0, \vec{P}^2) \,=\, \Lambda^2 \left(e^{P_0/\Lambda} + e^{-P_0/\Lambda} - 2\right) - e^{P_0/\Lambda} \vec{P}^2\,.
\ee

In order to determine the boost transformation of a system of two particles with momenta $(p,q)$ and a total momentum $p\oplus q$ we introduce the notation
\be
J^{0i}_{(2)} \,=\, x^\mu \psi_{L\,\mu}^{0i}(p, q) + y^\mu \psi^{0i}_{R\,\mu}(p,q)
\ee
for the boost generator acting on a two-particle system written in terms of the canonical phase space coordinates $(x, p)$, $(y, q)$. In correspondence with the choice $\varphi_L(p,q) = \varphi(p)$ for the implementation of locality, we make the choice $\psi_{L\,\mu}^{0i}(p, q) = \psi^{0i}_\mu(p)$ so that
\be
\lbrace p_\mu, J^{0i}_{(2)}\rbrace \,=\, \psi^{0i}_\mu(p)\,,
\label{delta(p)}
\ee
and the first momentum variable $p$ has the transformation under boost found previously in the case of a one particle system. For the second momentum variable we have
\be
\lbrace q_\mu, J^{0i}_{(2)}\rbrace \,=\, \psi^{0i}_{R\,\mu}(p,q)\,.
\label{delta(q)}
\ee
In order to determine $\psi^{0i}_{R\,\mu}(p,q)$ we impose the relativity principle as the requirement that the momentum of a Lorentz transformed one-particle system with a momentum $p\oplus q$ should be equal to the composition of the Lorentz transformed momentum variables of a two-particle system with momenta $(p, q)$. This requirement implies that
\be
\psi^{0i}_\mu(p\oplus q) \,=\, \frac{\partial(p\oplus q)_\mu}{\partial p_\nu} \psi^{0i}_\nu(p) + \frac{\partial(p\oplus q)_\mu}{\partial q_\nu} \psi^{0i}_{R\,\nu}(p,q) \,.
\label{RP}
\ee
When one inserts $\psi^{0i}_\mu(p)$ of Eq.~\eqref{psi(0i)}, and the composition law of Eq.~\eqref{mcl-bcb} derived from the implementation of locality, one can solve (\ref{RP}) for $\psi^{0i}_{R\,\mu}$. The result is
\be
\psi^{0i}_{R\,0}(p,q) \,=\, e^{-p_0/\Lambda} \,\psi^{0i}_0(q)\,, \qquad
\psi^{0i}_{R\,j}(p,q) \,=\, e^{-p_0/\Lambda} \,\psi^{0i}_j(q) + \frac{p_k}{\Lambda}\,\psi^{ik}_j(q)\,.
\label{psiR}
\ee
The boost transformation for the two momentum variables $(p, q)$ [Eqs.~\eqref{delta(p)}-\eqref{delta(q)}] is related with the coproduct of boosts of $\kappa$-Poincaré in the bicrossproduct basis~\cite{KowalskiGlikman:2002jr}
\be
\Delta(N_i) \,=\, N_i \otimes I \,+\, e^{-P_0/\Lambda} \otimes N_i + \frac{1}{\Lambda} \epsilon_{ijk}P_j \otimes M_k\,,
\label{coboost}
\ee
where $M_k$ are the rotation generators. From Eq.~\eqref{coboost}, one has
\be
\left[P_\mu \otimes I, \Delta(N_i) \right] \,=\, \left[P_\mu, N_i\right] \otimes I\,,
\ee
to be compared with Eq.~\eqref{delta(p)}, and
\be
\left[I \otimes P_\mu, \Delta(N_i)\right] \,=\, e^{-P_0/\Lambda}  \otimes \left[P_\mu, N_i\right] + \frac{1}{\Lambda}\epsilon_{ijk} P_j \otimes \left[P_\mu,M_k\right]\,,
\ee
to be compared with Eqs.~\eqref{delta(q)} and~\eqref{psiR}.

\end{document}